\documentclass[]{iopart}
\pdfoutput=1
\usepackage[english]{babel} 
\usepackage[utf8]{inputenc}

\usepackage[T1]{fontenc}

\usepackage[pdftex]{graphicx}  
\usepackage{dcolumn}   
\usepackage{bm}        
\usepackage{amssymb}   
\usepackage{cite}
\usepackage{subfigure}

\usepackage{subfigure}
\usepackage{hyperref}
 \expandafter\let\csname equation*\endcsname\relax
  \expandafter\let\csname endequation*\endcsname\relax
\usepackage{amsmath}

\begin{document}

\title[Lognormality and oscillations in RNAseq towards gene ends.]{Lognormality and oscillations in the coverage of high-throughput transcriptomic data towards gene ends.}

\author{Nicolas Innocenti$^1$ and Erik Aurell$^{1,2}$}
\address{$^1$ Department of Computational Biology, KTH Royal Institute of Technology, AlbaNova University Center, SE-10691 Stockholm, Sweden}
\address{$^2$ Aalto University, Department of Information and Computer Science, PO Box 15400, FI-00076 Aalto, Finland}
\ead{\mailto{njain@kth.se} and \mailto{eaurell@kth.se}}

\date{\today}

\begin{abstract}
High-throughput transcriptomics experiments have reached the stage where the count of the number of reads
alignable to a given position can be treated as an almost-continuous signal. This allows to ask questions of 
biophysical/biotechnical nature, but which may still have biological implications. Here we show that
when sequencing RNA fragments from one end, as it is the case on most platforms, an oscillation in the
read count is observed at the other end. We further show that these oscillations can
be well described by Kolmogorov's 1941 broken stick model. We investigate how the model can
be used to improve predictions of gene ends (3' transcript ends) but conclude that with present
data the improvement is only marginal. The results highlight subtle effects in
high-throughput transcriptomics experiments which do not have a biological origin, but which
may still be used to obtain biological information. 
\end{abstract}
\maketitle

\section{Introduction}
Next Generation Sequencing (NGS) is the common name used for most sequencing platforms currently
(August 2013) available on the market and widely used in genomic and post-genomic research.
In the present paper we investigate a common effect on most such platforms which is
relevant in transcriptomics \textit{i.e.} when they are used to sequence 
RNA extracted from some organism. Let us first remark that the outcome of an NGS experiment is 
a number count for each position on the genome of the organism under investigation.
This number count is called the \textit{coverage} and the average of the coverage over
the genome, or over the expressed parts of a genome in a transcriptomics experiment, is
called the \textit{sequencing depth}. For the sequencing of the human genome a sequencing
depth of 8-9x was used \cite{hgp}, i.e. the total length of all usable sequenced fragments was equivalent to 8-9 times the length of the studied genome. With current technologies one can
go much higher. For the bacterial transcriptomics data which will be used in the present
study the sequencing depth was about $450$x and we can therefore treat our coverage
signal as almost continuous. In a larger (more costly) experiment today, or in the future,
one can certainly also envisage such sequencing depths on human data which opens up the
interesting perspective of using data analysis techniques from physics or signal processing
in genomics more broadly.

All the common NGS platforms currently available on the market 
function by sequencing in parallel many short DNA molecules, typically 25-500 nucleotides (nt) in length \cite{Metzker:2010fk}. 
These short molecules are obtained by fragmenting DNA/RNA molecules using either 
physical or biochemical means such as nebulisation, sonication or random enzymatic digestion \cite{10.1371/journal.pone.0028240}. The result of sequencing one short molecule is called a read. 
In this work we will use data from the SOLiD platform where fragmentation and size selection are performed targeting fragments significantly longer than the read length \cite{Solidmanual,10.1371/journal.pone.0018595,SeanKennedy}. Consequently, only the first nucleotides of each fragment (5'-end) are read by the machine leading to a systematic truncation of the fragment ends (3'-ends).  This effect has often been ignored based on the assumption that the bias averages out for a 
large number of fragments coming from multiple identical molecules. The inspiration for this work is the observation
that these effects in fact do \textit{not} average out, but on the contrary lead to observable oscillations in the read coverage signal close to 3'-ends of RNA transcripts.

Perhaps surprisingly we were able to describe those
effects, on the average, by Kolmogorov's 1941 broken stick model~\cite{brokenstick},
which was, according to authorative sources~\cite{Frischbook}, one of the inspirations
for Kolmorov's 1962 refined theory of turbulence. The model can be used to improve
predictions of ends of transcripts (3'-ends) albeit we can only show a marginal
improvement on a gene-per-gene basis using current data. The reasons for 
this, which arise from other problems or
``artefacts'' in the sequencing, are discussed below.
 
The paper is structured as follows: in \sref{sec:theory} we recall briefly the broken stick model
and show how oscillations near 3'-ends can be generated by such a process.
In \sref{sec:data} we discuss our data, and in \sref{section:3primeend}
we discuss how the theory combined with such data can be used to predict gene 3'-ends.
In \sref{sec:results} we show that the predicted oscillations are well displayed on
average but that the model only marginally improves the prediction of single gene 3'-ends, for
reasons unrelated to the present work, which we discuss.
In \sref{sec:conclusion} we summarize and discuss our results.

\section{Reads and coverage}
\label{sec:theory}
The background to the theory is that biological RNA molecules very often are too long
to be directly handled by the sequencing platform. For instance, a typical bacterial gene
may have transcripts of length 1000 nucleotides or longer, while an eukaryotic 
transcript can be many times that length. Biologists have many ways to divide up a
long RNA (or DNA) molecule into shorter fragments, but one of the main one is
\textit{sonication}, or shaking by ultrasound until the molecule breaks.
An unbiased RNA fragmentation can then be seen as a sequential random process that, for $N$ fragments at a step $t$ creates $N+1$ fragments at step $t+1$ by selecting an existing fragment independently of its length and breaking it in two not necessarily equal pieces. Such a process is a particular case of Kolmogorov's Broken Stick model, and leads to fragments with lengths following a log-normal distribution  \cite{brokenstick,brokenstick2} with probability distribution function given by 
\begin{equation}
P(L)=\frac{1}{L\sqrt{2\pi s^2}}\, e^{-\frac{\left(\ln L- m\right)^2}{2 s^2}}, \label{eq:lognormal}
\end{equation}
where $L$ is the fragment length and the parameters $m$ and $s$ are given by

\begin{align}
s  &= \sqrt{\ln \left(   \frac{\sigma^4}{L_0^2}-1 \right) }, &  m = \ln \left( \frac {L_0^2}{\sqrt{\sigma^4 + L_0^2}} \right),
\end{align}
where $L_0$ is the average and $\sigma^2$ the variance of the fragment length distribution.

The effect we are interested in arises from each of these fragments being read from one end
(the 5'-end). In particular, the very last fragment which extends all the way to the gene end (3'-end)
is read from its other end (its 5'-end). Therefore, unless being shorter than
the number of nucleotides read by the platform, the last fragment will typically not be read all the way
to the end of the gene. This means that the \textit{coverage signal}, the number of reads
which can be obtained from a given genomic position, will tend to diminish towards the end of the
gene, and at the very end the signal will be low. This leads to considerable error if one tries
to use the signal to locate the 3'-end precisely. 

However, a more subtle effect also gives rise to oscillations in the coverage signal. Consider 
the situation where there are preferred lengths $l''$ and $l'$ of the last two fragments. If so,
the sequencing platform will deliver more than average number of reads that align to genomic
positions respectively $l'$ and $l'+l''$ nucleotides before the gene end, while, in addition to
the depletion at the 3'-end, there will also be less than average number of reads aligning to
positions between $l'$ and $l'+l''$. In other words, the coverage signal will display oscillations
towards the gene end. 

In reality one would observe a distribution of values of $l''$ and $l'$ and the
oscillations will only show up on the average.
We now suppose that these distributions can be modeled using the broken stick model. 
We set up a Monte Carlo scheme that creates fragments with a distribution given by \eref{eq:lognormal} from the last $M$ nucleotides of numerous identical copies of an RNA molecule and build  $c_k(L_0,\sigma)$, the coverage $k$ nucleotides upstream of the transcript 3'-end that one expects to obtain provided that only the first $N$ base pairs of each fragment are sequenced, i.e.
\begin{equation}
c_k(L_0,\sigma) = A \sum_{i\in I_{L_0,\sigma}} \mathbf{1}_N(X_i-k),\ \ \ \ k=1,2,...M,\label{eq:montecarloscheme}
\end{equation}
where $X_i$ corresponds to the distance from the first nucleotide of a fragment $i$ to the 3'-end on the original RNA molecule, $A$ is an arbitrary constant and $\mathbf{1}_N(x)$ is an indicator function equal to one on $[0,N]$ and zero elsewhere.  Each fragment $i$ is taken from a set $I_{L_0,\sigma}$ obtained by generating fragments with lengths distributed according to \eref{eq:lognormal} and excluding the ones shorter than the read length $N$. 

This Monte Carlo scheme gives us the coverage signal ``as it should be''.
Provided that the parameters $L_0$ and $\sigma$ are known for a given experiment and that the existence of a 3'-end is known within a genomic region $D$ of reasonable size, its precise location $z$ can then be obtained by maximising the overlap between the predicted pattern and the experimental coverage, i.e. by solving \begin{equation}\operatorname*{arg\,max}_{z \in D} \sum_{k=0}^M   c_k(L_0,\sigma) ~\tilde c_{k,z}, \label{eq:predictor} \end{equation} where  $\tilde c_{k,z}$ denotes the coverage obtained from the experiments $k$ nucleotides upstream of a genomic position $z$.

\section{High-throughput RNA sequencing of the human pathogen E faecalis}
\label{sec:data}
The data which we used were obtained as part of an ongoing effort to understand
genes, their regulation and generally new biology in the human pathogen
\textit{Enteroccocus faecalis}~\cite{Fouquier2011}. This bacterium, which is commonly found
in the lower parts of the human digestive apparatus, is also an
opportunistic pathogen and ranks approximately fourth or fifth as
a cause of nosocomial infections, or "hospital disease'', world-wide \cite{Palmer29102010}.
It is especially medically important as it easily acquires antibiotic resistance,
potentially leading to severe disease outcomes in a hospital environment.
It is also interesting from a general scientific point of view as an example
of an organism which has been known for a long time -- and is likely much more important to human
health than \textit{e.g.} \textit{Escherichia coli} -- but which is still comparatively
much less well known.

We proceed to outline what is meant by the last phrase in the preceeding paragraph by following what a bioinformatician (or a physicist using bioinformatic tools) would do. First, we would like to know which are the genes in the organism and what could their function be. If the organism is sequenced, i.e. if the genome is available and deposited in a public database such as Genbank, there are standard tools which will take the genome as input and will give a (hypothetical) gene list as output.

The reference variant of E. faecalis is the strain v583, which is resistant to the antibiotic vancomycin, and was isolated in Barnes Hospital, St. Louis, Mo., USA in 1987 \cite{Sahm01091989}. Standard tools to annotate bacterial genomes such as Glimmer  \cite{Delcher01121999} will give a gene list where a putative function of each putative gene is given by analogy with similar functions of genes coding for a similar protein in other bacteria, in fact mainly in E. coli. Glimmer or other tools give in fact only what are known as open reading frames (ORFs), which are DNA sequences that may code for a protein, but they do not give the full gene from the start (the 5'-end) to end (the 3'-end). It has long been recognised that the residues in the flanking regions of DNA at the 5' and 3'ends, which do not code for amino acids are crucial for regulation of transcription \cite{Ptashne}. Unfortunately, determining 5'-ends of regulatory elements automatically from a sequence is more difficult than determining an ORF and cannot be done reliably. Hence the interest to deduce such genomic elements by other means.

We analysed a total of 6 datasets from RNA sequencing (RNA-seq)  experiments performed on \textit{Enteroccocus faecalis} strain v583. 
Four of those originate from a single RNA extraction sequenced on the SOLiD v3 platform using the plain single stranded RNA sequencing protocol from the platform manufacturer. In experiments labeled here and below S3sr1A and S3sr1B, all RNA was sequenced while in S3sr0A and S3sr0B the ribosomal RNA (rRNA) was removed beforehand using Ambion MICROBExpress Bacterial mRNA Enrichment Kit. We note that in bacteria rRNA may constitute as much as 90\% of total RNA in an extract, and that removing rRNA before analysis is a common practice. From the point of view of the present work, this difference is however not important and the experiments with or without rRNA removal can therefore be considered as replicates. The different suffixes A or B correspond to independent RNA-seq performed on two samples from the same RNA extract separated after fragmentation, i.e. they are technical replicates of the experiments.  

The two further datasets labelled  S55sr1 and S55rr1 correspond to single strand RNA-seq on the SOLiD 5500 platform from two RNA extractions obtained from bacteria grown in different growth conditions. Ahead of manufacturer's protocol, short RNA oligos were ligated to the 5'-ends as described in~\cite{Fouquier2011}.
 
 An important aspect of the preparation protocols used in all the experiments introduced above is that RNA (or DNA) molecules are fragmented and size selected targeting a length greater than 100 nt \cite{Solidmanual,10.1371/journal.pone.0018595,SeanKennedy}. These fragments are then amplified and attached to beads in an emulsion polymerase chain reaction (emulsion PCR) before being sequenced. The sequencing itself is performed on the 50 first nucleotides --- or more precisely on the first 50 nucleotide \emph{transitions}, as obtained with SOLiD 2-base color encoding \cite{solidcolorspace} --- of each fragment, leaving the remaining part of the molecule unread. 

Every one of the above introduced datasets contains 60 to 100 million reads, each with a length corresponding to 50 color encoded nucleotide transitions, each position of each read being accompanied with a quality score indicating how well this particular position was read by the sequencer. 

Once you have a read you have to find where on the genome it came from. This procedure is called alignment and while it could often be done directly with high accuracy, it is better practice to use a pre-existing bioinformatic tool which avoids known pitfalls and which will indicate how likely is the information given to be correct. We aligned our data using Bowtie (version 0.12.7) \cite{bowtie}, an open source alignment software widely used in RNA-seq data analysis. In the configuration we used, the software reports an alignment if the read matches the genome with less than 2 errors in the 28 first nucleotides of the read and if the sum of the read qualities at all mismatching positions is lower than a threshold. If multiple positions satisfy those conditions, all of them are reported as valid alignments.  After alignment, we calculate the coverage and take multiply mapped reads into account by dividing their contribution to the read count by the number of matches.  Due to the specificities of the aligner, reads are converted to 48 nt long fragments during the alignment, thus reducing the practical read length to this number.

 For every dataset, we were able to align between 40 and 60\% of reads to the genome, which corresponds to an average coverage over 450x of the 3.2 million nucleotides present in the E. faecalis v583 genome.\\ 

\begin{figure}
\centering
\includegraphics[width=.75\textwidth]{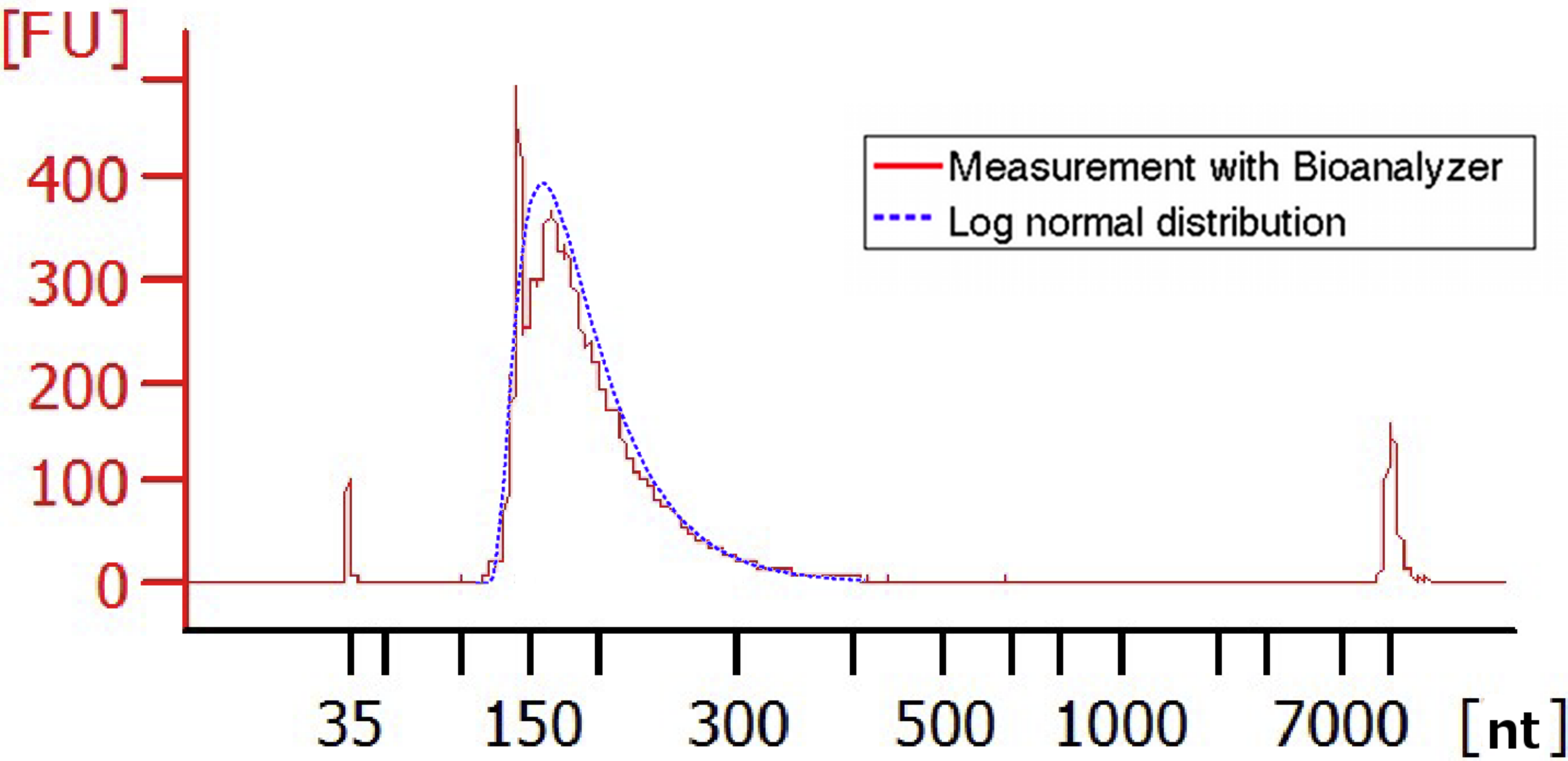}
\caption{Length distribution of an RNA sample measured with an Agilent 2100 Bioanalyzer after fragmentation and library preparation adding adapters with total constant length of 128 nt. The dashed blue line shows a log-normal distribution with mean 65 and standard deviation 60 shifted by 128 nt to the right. The peak observed in the measurement at 141 nt corresponds to an excess of synthetic short RNA sequences artificially added to the particular sample used.\label{fig:fragmentlength}}
\end{figure}

In order to confirm that RNA fragmentation results into the postulated log-normal distribution, we measured the fragment length distribution for another RNA sample prepared for future RNA-seq using an Agilent 2100 Bioanalyzer (\fref{fig:fragmentlength}). The match with a log-normal distribution is clear except for a peak corresponding to fragments of 13 nt, which corresponds to synthetic RNAs artificially added to this particular sample. Unfortunately, no measurements of this type are available for the samples used to generate our 6 datasets. 

\section{Transcript 3'-ends}
\label{section:3primeend}
Bacteria have two major mechanisms for transcription termination, the rho-dependent terminator and the rho-independent,  or intrinsic, terminator.  In E. coli, each of the mechanism accounts for the termination of about half of the genes, the rho-dependent termination playing a major role in transcription regulation \cite{Peters2011793}.

In the rho-independent termination, due to nucleotide-pairing, the transcribed RNA folds into a stem-loop structure of 7-20 nt in length, rich in cytosine-guanine nucleotide pairs and followed by a poly-uracil (poly-U) tail. This stem-loop binds to the RNA-polymerase complex, which catalyses the synthesis of RNA from the DNA template, causing a pause in the transcription while the weak bonds between the poly-U tail and the adenines on the DNA allow for the complex to detach from the DNA strand \cite{Wilson12091995}. Such terminators can be quickly and reliable predicted using TranstermHP \cite{transtermHP},  a now standard computational tool that searches whole genomes for occurrences of sequences that could lead to the terminators described above by systematically inspecting the folded structure of RNA transcripts ahead of all adenine-rich regions of the genome that could be coding for suitable poly-U tails. The software takes as input a whole genome and outputs a list of potential rho-independent terminators described by the beginning and end of the stem-loop as well as a confidence level corresponding to the probability that the candidate actually is a functional terminator. 

In the rho-dependent termination, the mechanism is mediated by a protein called \emph{Rho factor}. Rho binds the freshly transcribed RNA in weakly folded regions at specific locations called rho utilisation (rut) site. Such sites are about 70 nt in length and cytosine rich, but no consensus sequence for a rut is known.  After binding, Rho moves downstream the RNA strand through an energy consuming mechanism. When approaching the region where RNA synthesis takes place, Rho unwinds the nascent RNA from the DNA strand and eventually terminates the transcription by causing the RNA polymerase to detach in a mechanism that is so far not well known \cite{Boudvillain2013118}. The transcription termination typically takes place between 10 to 100 nt downstream of the rut site, its exact location depending on the relative speed of translocation of Rho on the RNA strand and the one of RNA synthesis of the RNA polymerase. Due to the complexity of this mechanism, there is no way described in the literature to reliably predict rho-dependent termination sites. \\

We predicted rho-independent terminators on the chromosome of E. faecalis v583 (NCBI reference AE016830) using TranstermHP (v2.08, with default options). Out of a total of 1851 predictions, only the 1229  terminators found with 100\% confidence were retained.  

\begin{figure}[b]
\centering
\includegraphics[width=.7\textwidth]{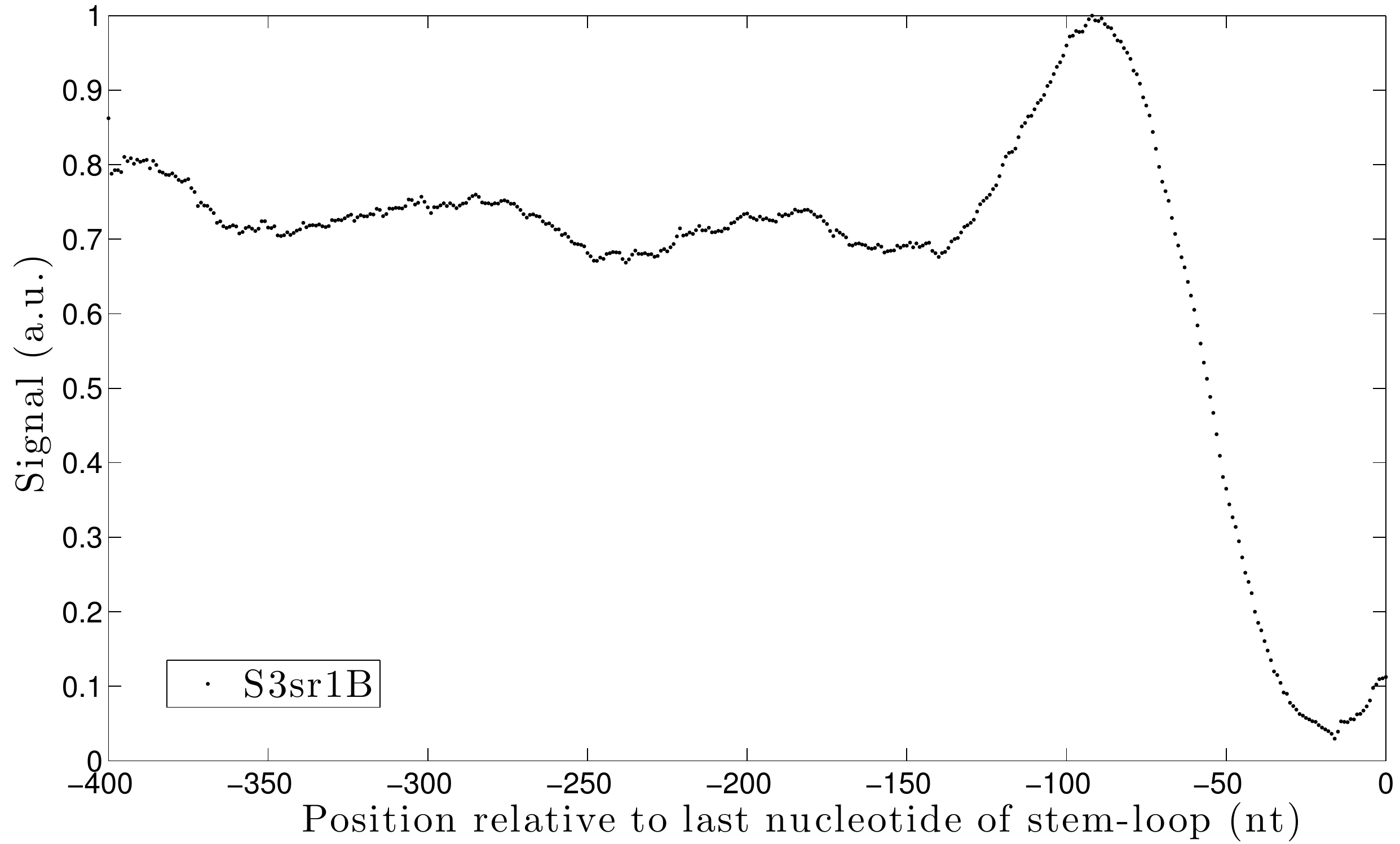}
\caption{Normalized coverage signal obtained from the S3sr1B dataset ahead of rho-independent terminator averaged over the 1229 rho-independent terminators predicted with 100\% confidence on E. faecalis chromosome using TranstermHP. \label{fig:signals}}
\end{figure}
Upstream of each selected terminator, we collected the read coverage obtained from RNA-seq on a window of 400 nt starting from the last nucleotide of the terminator stem-loop as predicted by TranstermHP. It is worth noting that the real 3'-end of the corresponding transcript is located downstream of this location, usually after a 4 to 9 nt long  poly-uracil (poly-U) tail \cite{Peters2011793,Carafa1990835}. The signal was normalised so that its integral over the region of interest is unity. Averaging over all the 1229 terminators, for the case of the S3sr1B dataset, lead to the signal presented in \fref{fig:signals}.

\begin{figure*}
\subfigure[~S3sr0A]{\includegraphics[width=.48\textwidth,trim= 1cm 0cm 2cm 0cm, clip]{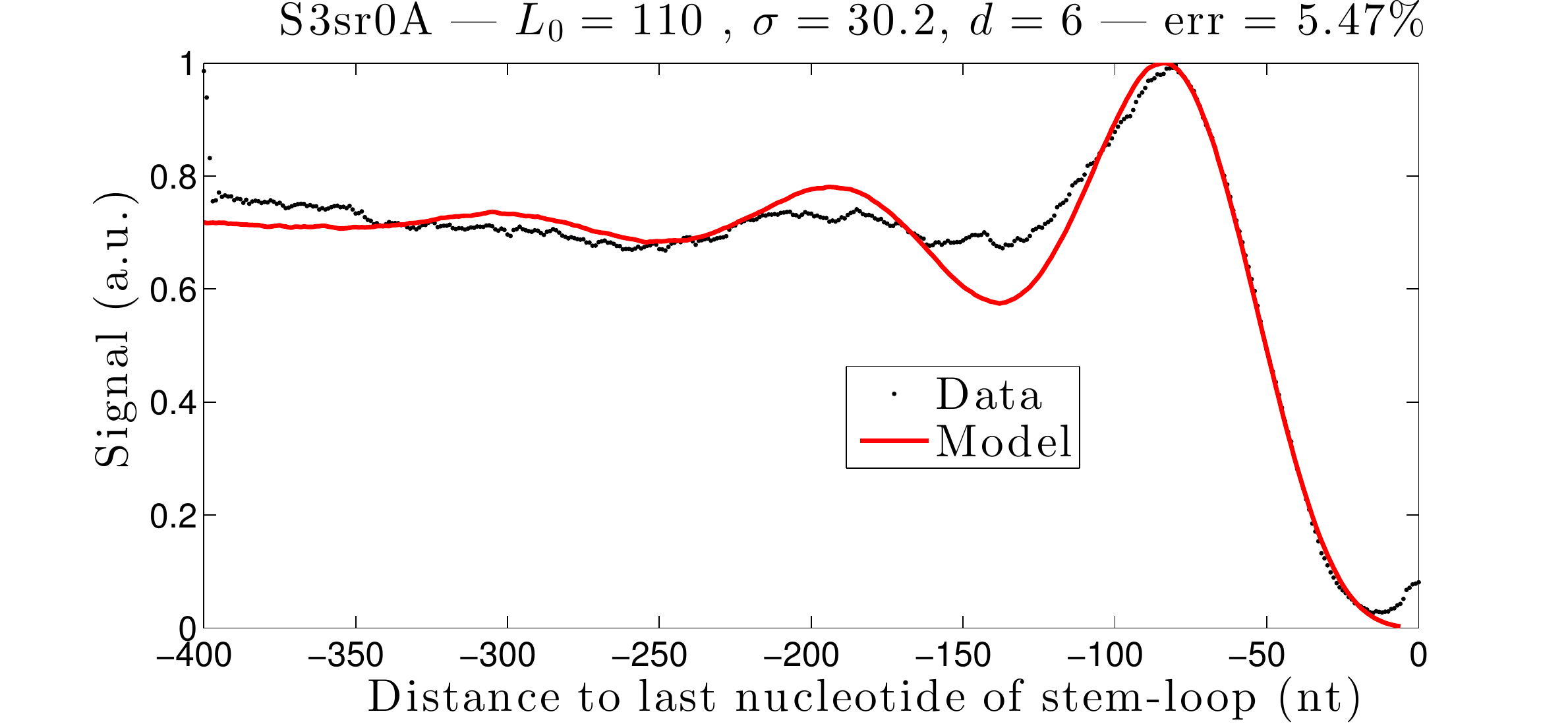}}\subfigure[~S3sr0B]{\includegraphics[width=.48\textwidth,trim= 1cm 0cm 2cm 0cm, clip]{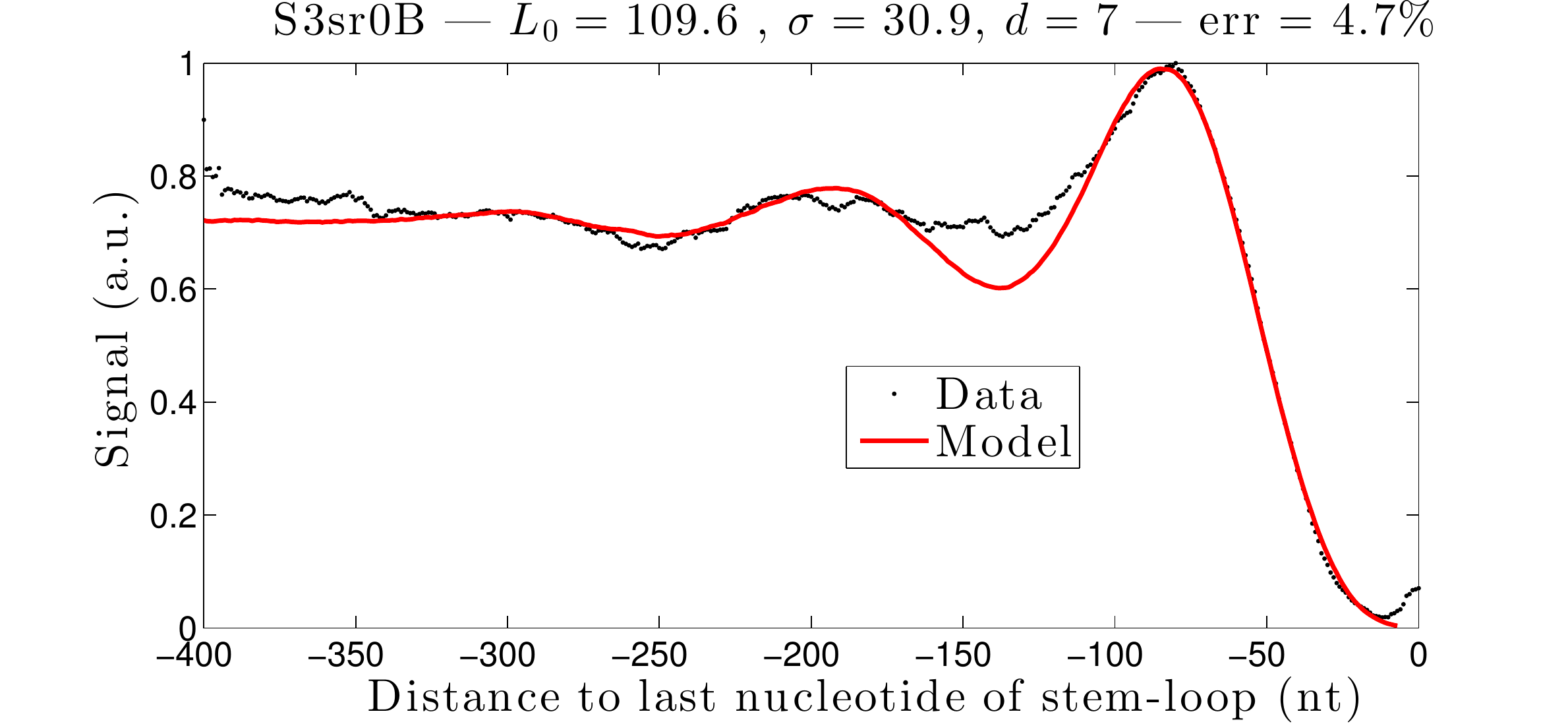}}
\subfigure[~S3sr1A]{\includegraphics[width=.48\textwidth,trim= 1cm 0cm 2cm 0cm, clip]{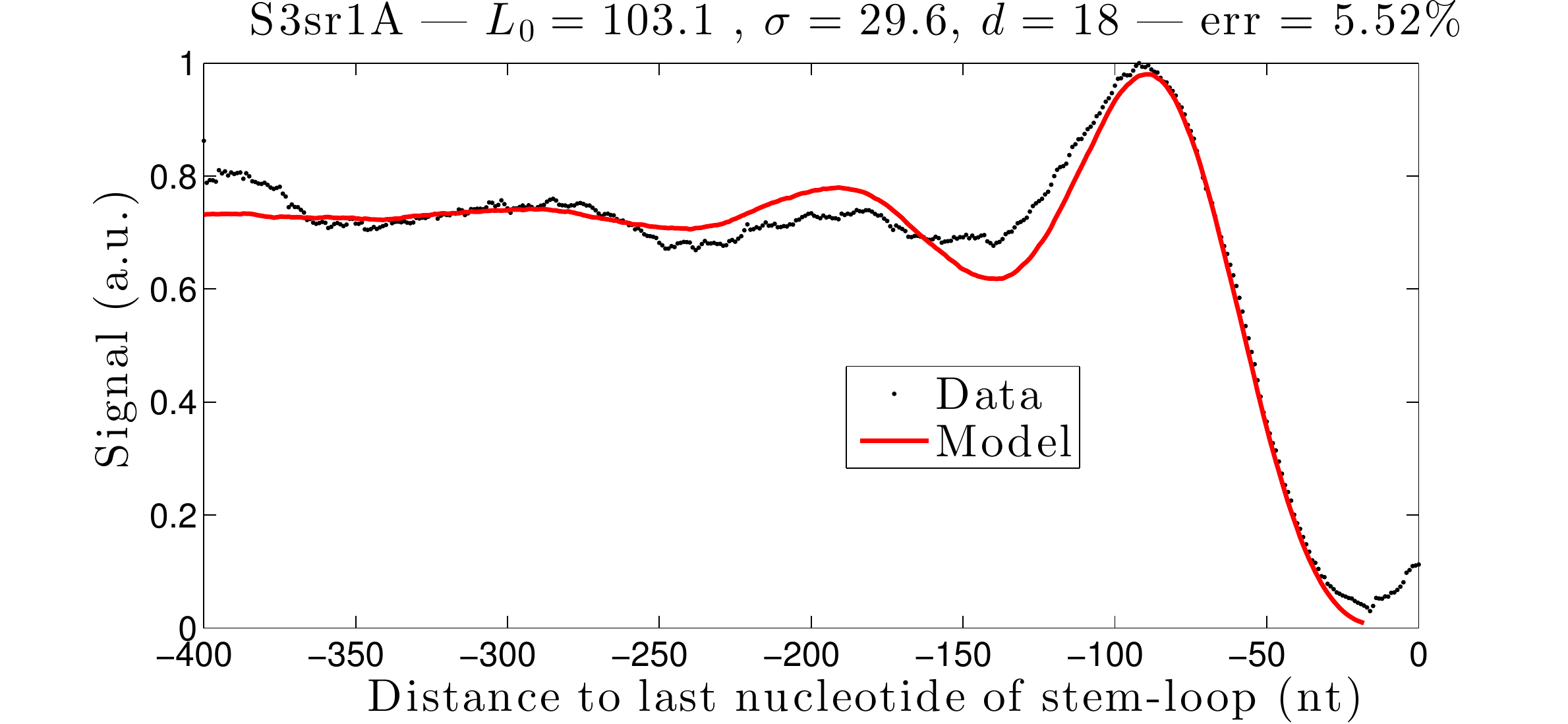}}\subfigure[~S3sr1B]{\includegraphics[width=.48\textwidth,trim= 1cm 0cm 2cm 0cm, clip]{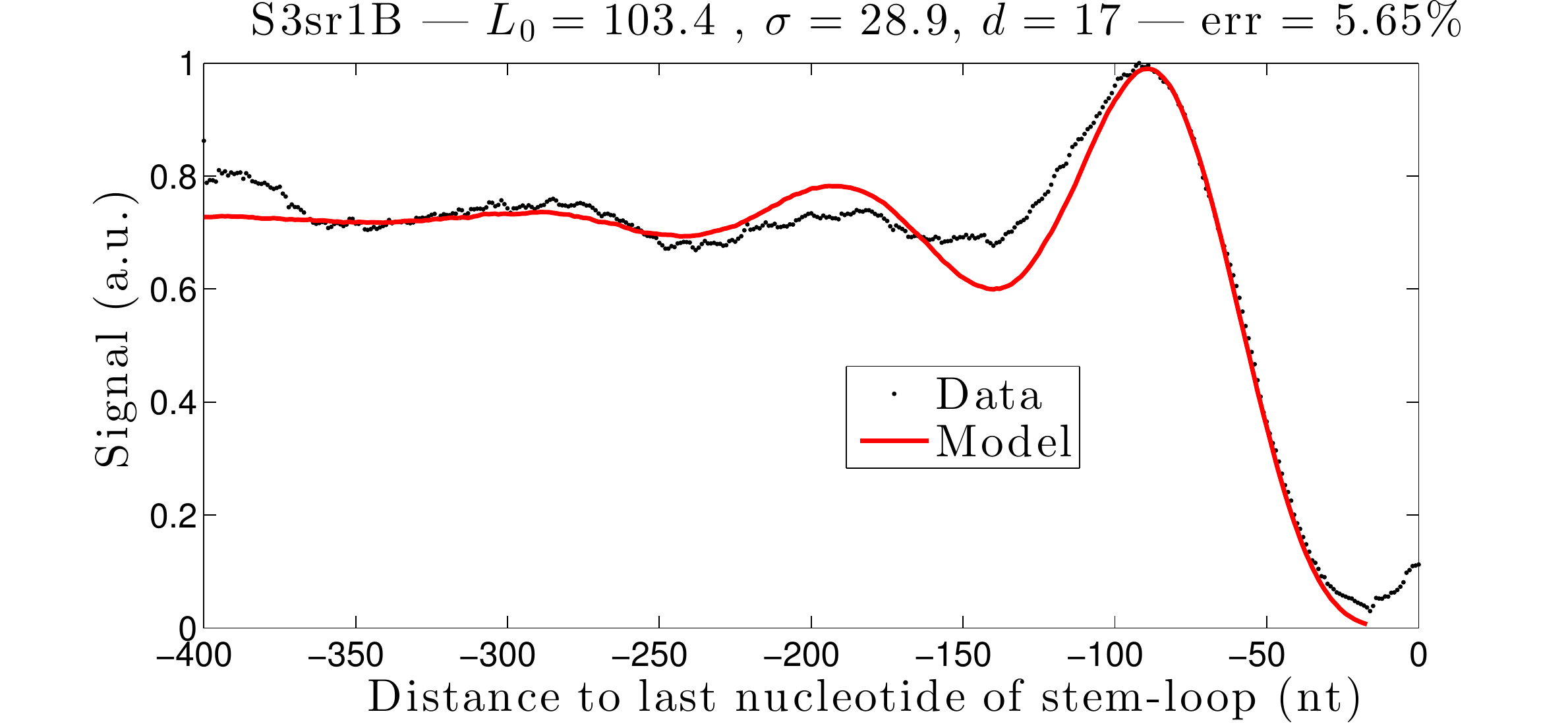}}
\subfigure[~S55rr1]{\includegraphics[width=.48\textwidth,trim= 1cm 0cm 2cm 0cm, clip]{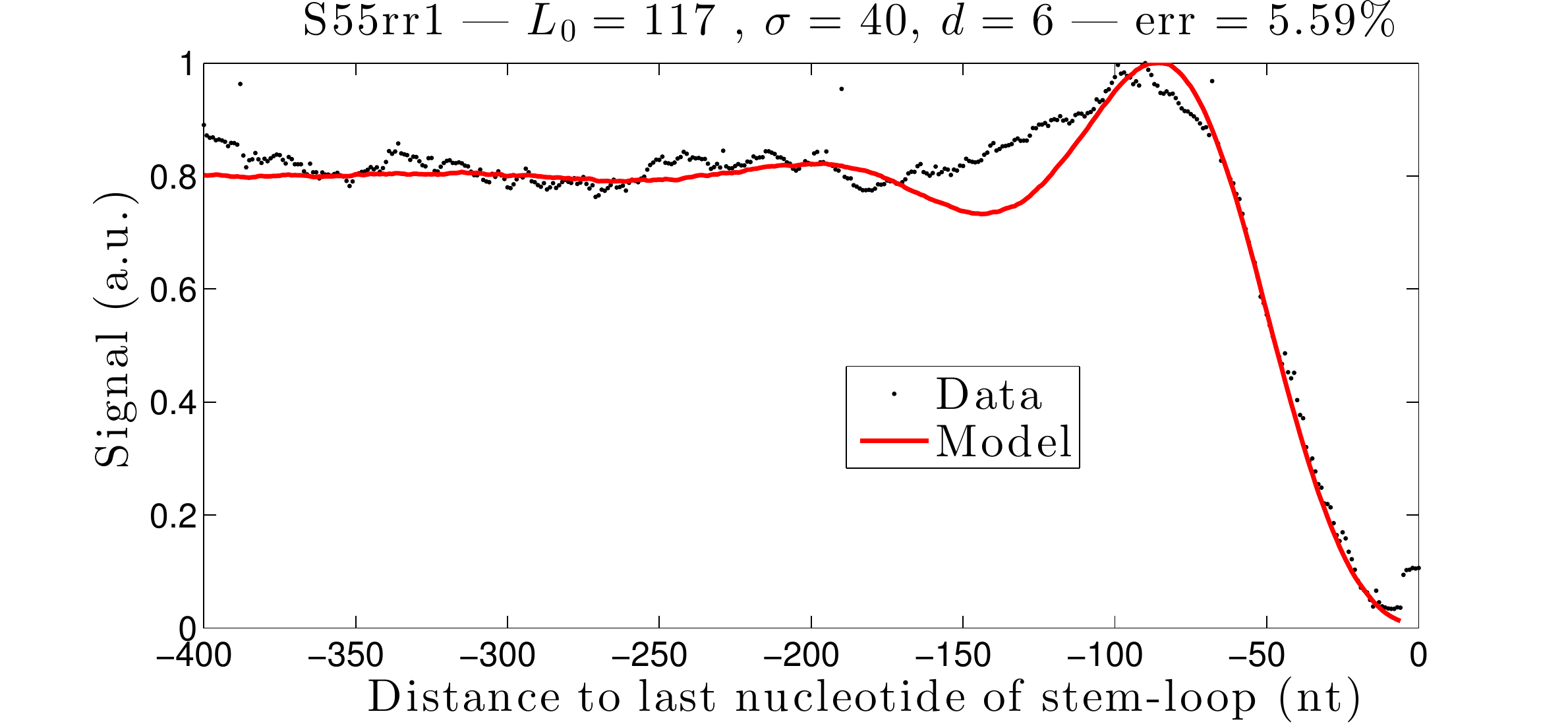}}\subfigure[~S55sr1]{\includegraphics[width=.48\textwidth,trim= 1cm 0cm 2cm 0cm, clip]{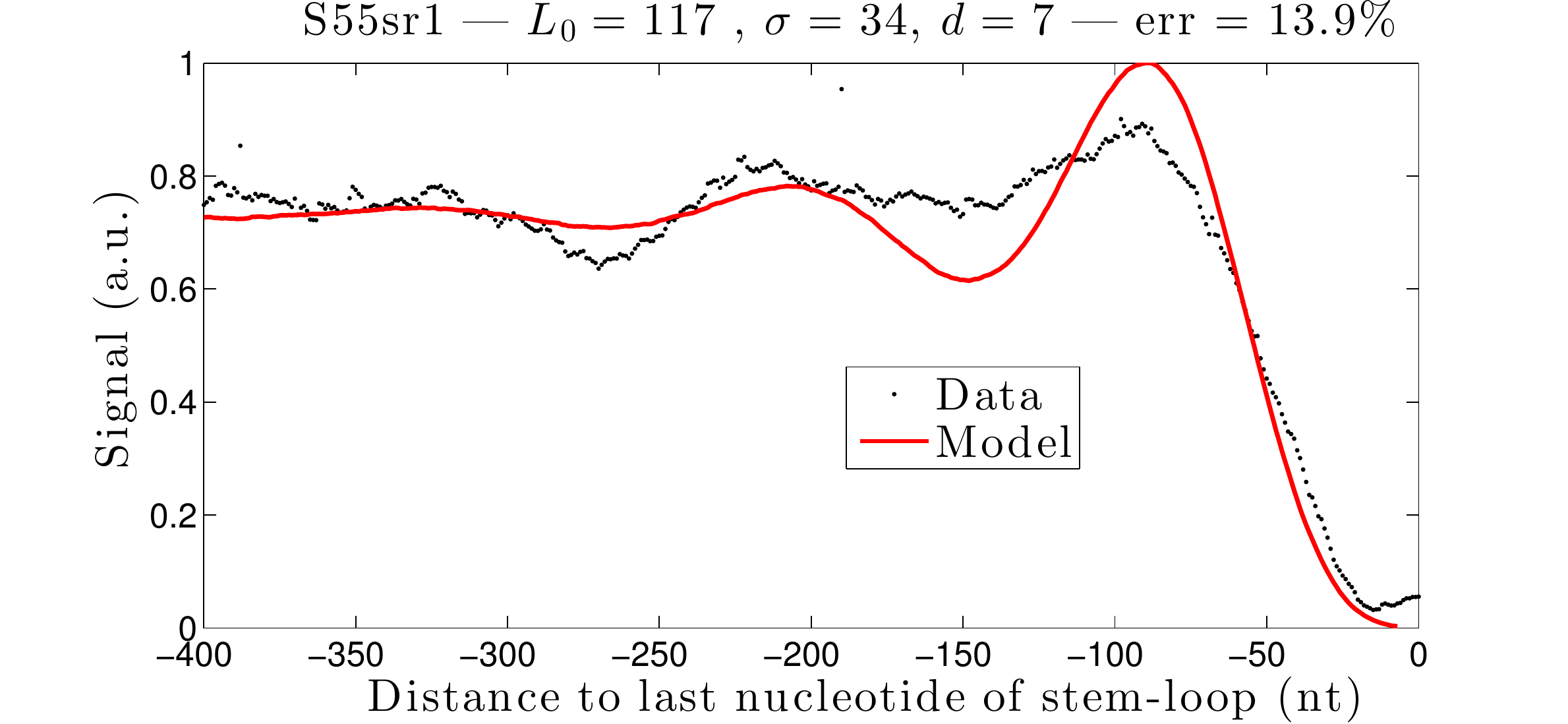}}
\caption{Comparison between the signals obtained as described in section \emph{Transcripts 3'-ends} from the 6 RNA-seq experiments and the ones predicted by the theory after fitting the parameters $L_0$, $\sigma$ and $d$. \label{fig:results}}
\end{figure*}

It can be easily seen that the signal falls to zero well before the location of the terminator and has a pronounced peak around position -80 nt. Upstream of the peak, one can notice 2 to 3 periods of a weak oscillation with a period on the order of 100 nt, with no plausible biological origin.  
\section{Results}
\label{sec:results}

We use  \eref{eq:montecarloscheme} from the previously described Monte Carlo scheme with a read length $N=48$ nt to generate the coverage expected from our theory. In every case, the scheme creates random fragments from $10^5$ copies of the original complete molecule, which appears to be sufficient for a relative  accuracy better than $10^{-2}$. Additionally, we introduce a parameter $d$ that corresponds to a positional shift representing the average distance between the last nucleotide of the stem-loop in the rho-independent terminator as predicted by TranstermHP and the real 3'-end of the corresponding molecule, i.e. the length of the poly-U tail following the stem-loop.  We fit $d$, the mean $L_0$ and the standard deviation $\sigma$ of the underlying log-normal distribution. We output the coverage at each position over 350 nt ahead of the last nucleotide of the stem-loop.  We measure the error $\varepsilon$ between the theoretical coverage $c$ obtained using the fitted  parameters and  $\tilde c$ , the one obtained from the data, as 
\begin{equation}
\varepsilon = {\sum\limits_{i=1}^{350} \left( c_{i+d} -  \tilde c_i \right)^2 } \Bigg{/}{~~~\sum\limits_{i=1}^{350} \tilde c_i^2 } ~~.
\end{equation}

Fitting the two datasets from the S3sr0 type (\fref{fig:results}(a) and (b)) results in an average fragment length of 110 nt and an estimated poly-U tail length $d$ of 6 nt, which is in  good agreement with what is commonly expected for this tail \cite{Peters2011793,Carafa1990835}. On the other hand, the datasets of the S3sr1 series (\fref{fig:results}(c) and (d))  result in a shorter estimated average fragment length and a poly-U tail length of 17 nt, which seems unrealistically high. As there is no reason for the two pairs of experiments to differ in the length of the poly-U tail (the two samples were prepared using the same protocol from the same RNA extraction and differ only by the removal of rRNA), we attribute this difference to an issue with data fitting and note that underestimating $L_0$ causes an overestimate of $d$. 

Furthermore, we note that the results for the 'A' and 'B' samples of  S3sr0 and S3sr1 are very similar to each other in every aspect. Since the members of each pair were separated after the RNA fragmentation step, this indicates a good reproducibility of the sample preparation and sequencing, and implies that the observed effects depend only on the steps preceding and including the fragmentation. 

The further datasets of the S55 series lead to much worse fits (\fref{fig:results}(e) and (f)) where only the right-most part of the signal corresponding to the fall before the terminator can be reproduced. Overall, the signal obtained from the data seems much more noisy and the pattern is less pronounced than in the examples previously described. This effect is likely related to the different protocol used in those experiments: the inclusion of tags at the beginning of RNA molecules modifies their lengths in a non controlled way, causing a blurring of the pattern.  Nevertheless, the parameters obtained from the fit are realistic and in line with those of the other four datasets. 

\begin{figure}
\centering
\includegraphics[width=.75\columnwidth,trim=0cm 0cm 0cm 0cm, clip=true,]{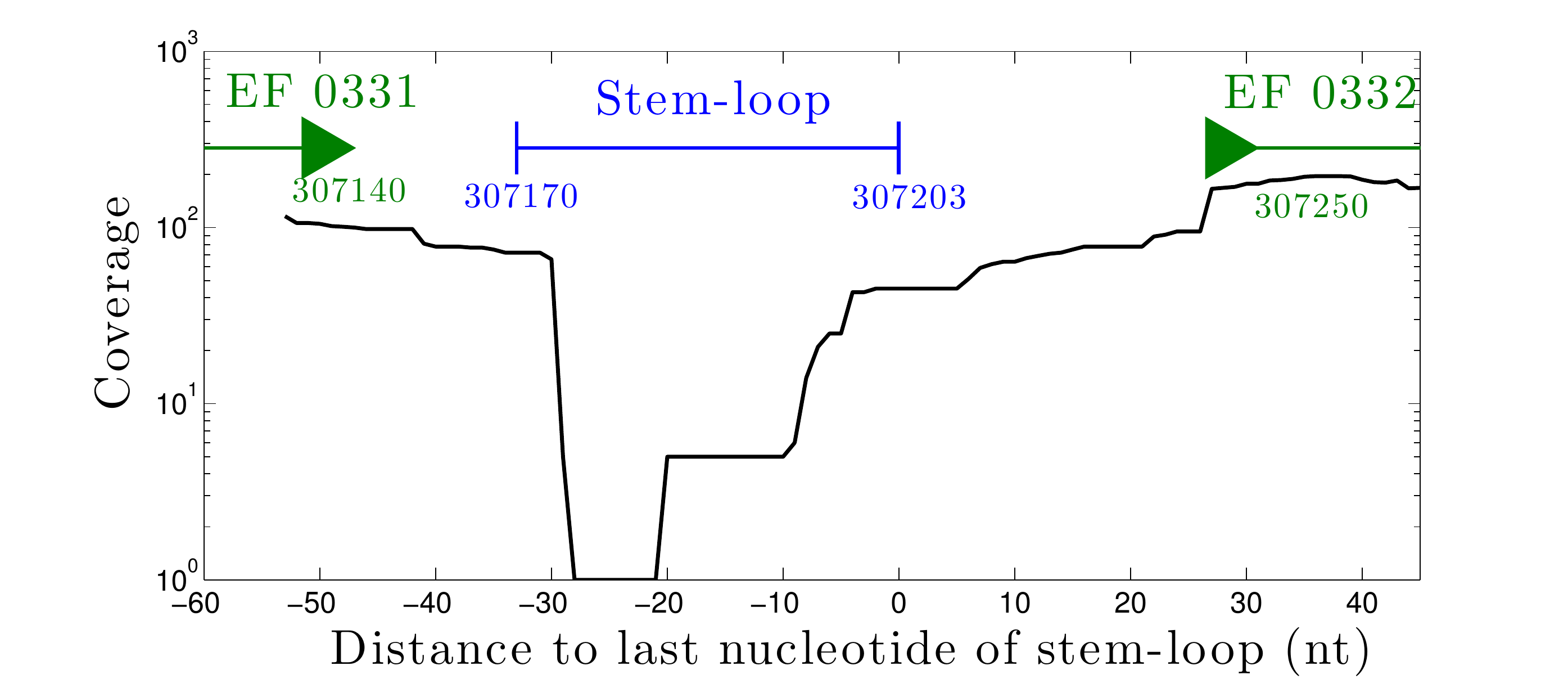}
\caption{Coverage signal from the dataset S3sr0A between genes EF 0331 and EF 0332 where the transcription of EF 0332 seems to start within the terminator of the previous gene. The end of the ORF of EF 0331 and the beginning of the one of EF 0332 are indicated by the green horizontal lines and the green arrows indicate the transcription direction. The stem-loop of the predicted rho-independent terminator, located between position 307170 and 307203 on the chromosome, is marked in blue.  The horizontal axis show the relative distance to the last nucleotide of the stem-loop of the predicted terminator and the numbers inserted in the plot indicate the position on the E. faecalis v583 chromosome.  \label{fig:TSSinterminator}}
\end{figure}

Finally we note that in all cases the experimental signal increases close to the transcript end, starting 5 to 10 nt ahead of the last nucleotide of the stem loop, which is at odds with the proposed theory. We observe that in a few cases the next gene downstream on the DNA seems to have its transcription start site within the DNA region corresponding to the rho-independent termination (example on \fref{fig:TSSinterminator}), which is enough to explain the effect observed in the average signal. Whether such transcription start sites are biologically relevant or result from sequencing artefacts in our experiments is unclear. 

\begin{figure}
\centering
\includegraphics[width=.75\textwidth,trim=1cm 0cm 1cm 0cm, clip=true]{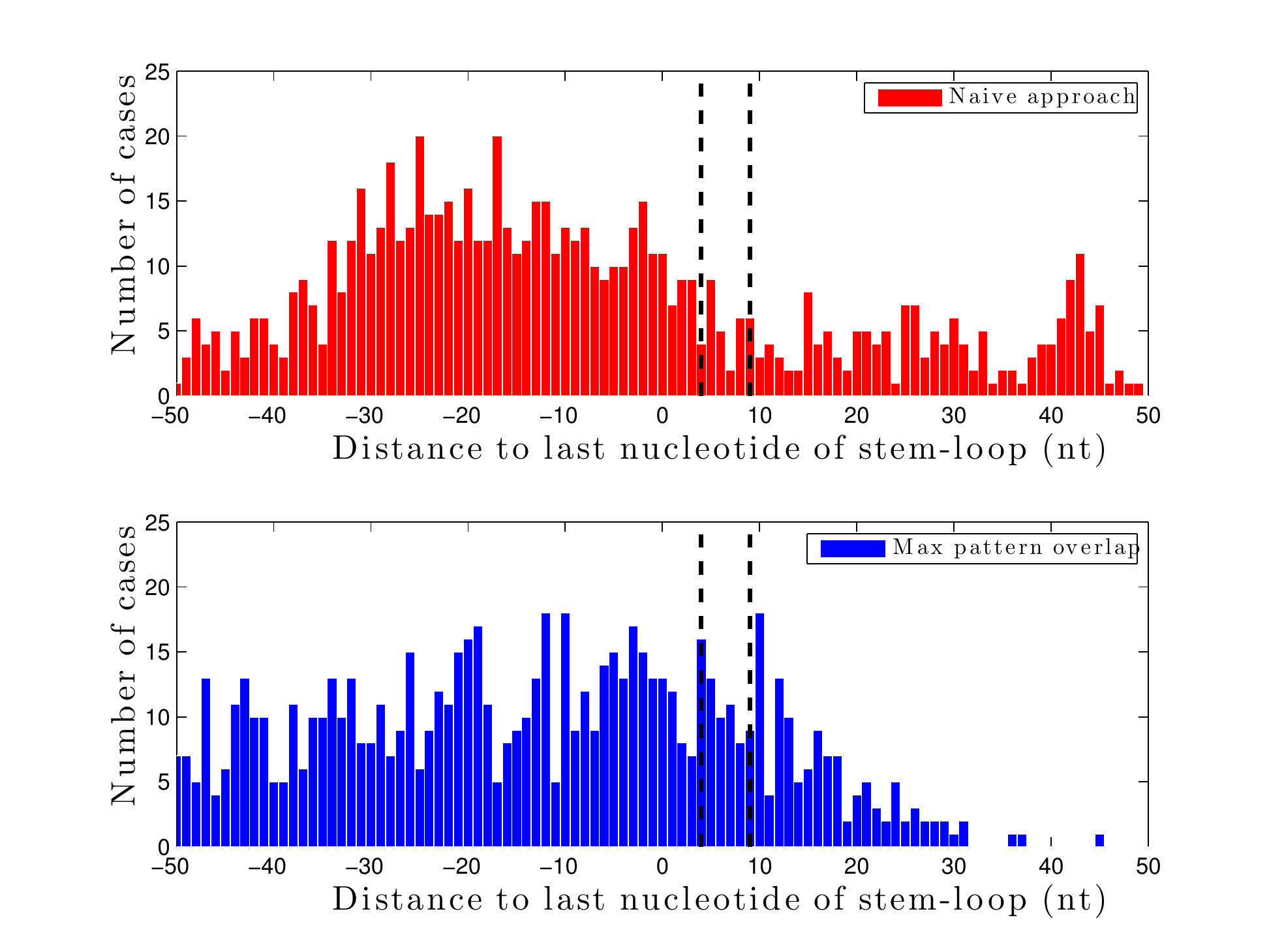}
\caption{Distribution of distances between the last nucleotide in the stem-loop of the rho-independent terminator and predictions of 3'-end of the corresponding transcript obtained either by a naïve approach searching for the closest locus where the signal goes to zero (top) or using the maximum pattern overlap method from \eref{eq:predictor} (bottom) for the 1229 terminators predicted by TranstermHP with 100\% confidence. The vertical dashed lines delimits the region between 4 and 9 nt downstream the terminator stem-loop where the actual 3'-end is expected. \label{fig:prediction}}
\end{figure}

\begin{table}[t]
\begin{tabular}{ | l | c | c |  c | c |}
\hline
 &  Nb. in [-50,50] & Nb. in [4,9] & Mean  & Std.  \\
 \hline
 Naïve approach & 745 & 32 & 24.83 & 13.41 \\
 Pattern overlap & 744 &  67 &  22.80   &     16.08    \\ 
 \hline
\end{tabular}
\caption{Summary of the data presented in \fref{fig:prediction} showing for both methods the number of predictions within certain ranges and the mean and standard deviation of the absolute distance between a prediction and the expected location of the 3'-end, assumed to be 7 nt downstream the last nucleotide in the stem-loop of the rho-independent terminator, as obtained from the fit in \fref{fig:results}(b) \label{tab:prediction}}
\end{table}

We now focus on the dataset S3sr1B that gives the lowest relative error and use our theory to predict individual 3'-ends near the 1229 predicted rho-independent terminators. We use as initial guess the location of the last nucleotide of the stem-loop and search within a region of 100 nt centered around this guess. We compare on \fref{fig:prediction} and \tref{tab:prediction} the predictions obtained by optimising \eref{eq:predictor} to a naïve approach that detects the locus closest to the initial guess where the coverage goes to zero. We notice that the number of predictions within a range of 100 nt around the initial guess is similar for both methods, while the number of predictions falling between 4 and 9 nt downstream of the last nucleotide of the terminator stem-loop (the location where the real 3'-end of the transcript is expected) is more than twice higher for the maximum pattern overlap method. We also note that the average absolute distance to the expected 3'-end is slightly improved by using our theory,  but its standard deviation is worse. Finally, we point out that the naïve approach predicts many 3'-ends far downstream the terminator (+30 nt and further) which are not biologically relevant. This effect is not present while using predictions based on our theory.

\begin{figure}
\centering
\subfigure[Region ahead of terminator at 739111 - 739129]{\includegraphics[width=.75\columnwidth,trim=2cm 0cm 3cm 0cm, clip=true,]{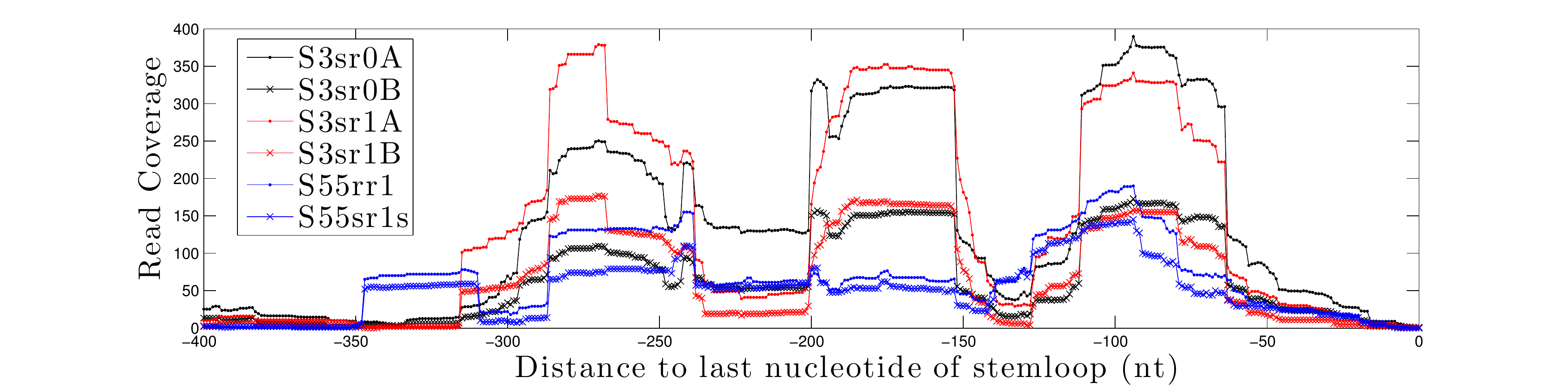}}
\subfigure[Region ahead of terminator at 273009 - 273026]{\includegraphics[width=.75\columnwidth,trim=2cm 0cm 3cm 0cm, clip=true,]{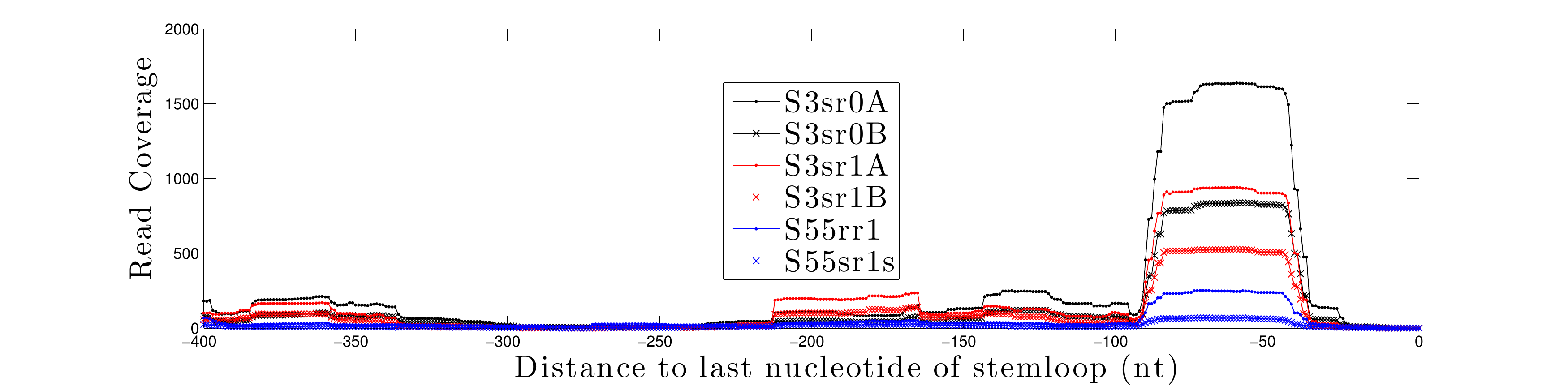}}
\subfigure[Region ahead of terminator at 595281 - 595333]{\includegraphics[width=.75\textwidth,trim=2cm 0cm 3cm 0cm, clip=true,]{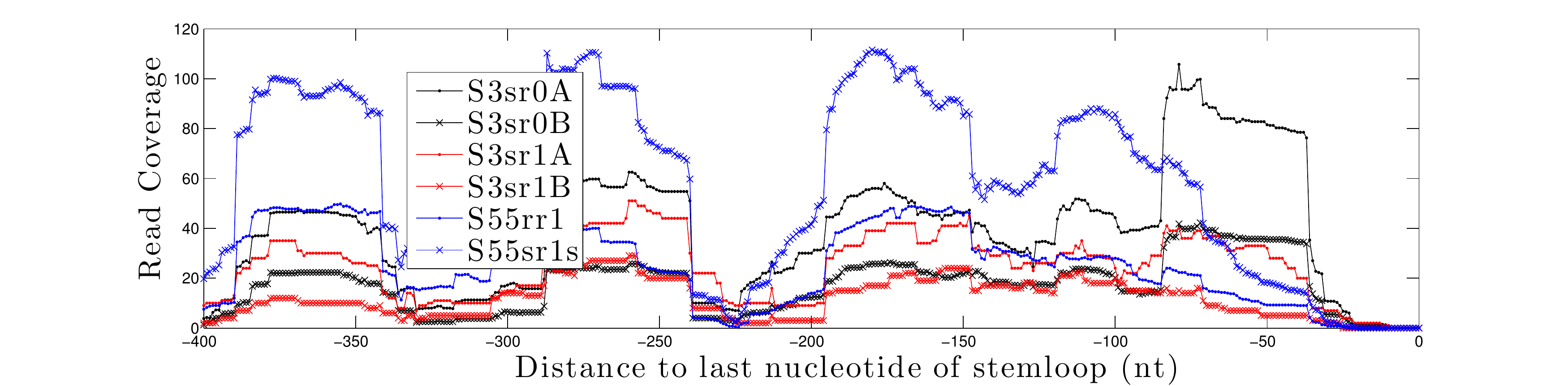}}
\caption{Examples of the read coverage ahead of predicted terminators showing patterns very different from the one predicted by the theory.\label{fig:sampleregions}}
\end{figure}

The reason for having only a relatively modest improvement is that, while we have shown that the proposed theory works well for the average signal ahead of gene ends, the pattern is not directly observed when considering the read coverage of one individual transcript. Often, the signal shows one or several box-shaped regions of high signal, with a position and periodicity only very roughly in agreement with the theory (\fref{fig:sampleregions}(a) and (b) ). In other situations, the pattern may be completely absent (\fref{fig:sampleregions}(c), in particular the blue curves).  Those effects are likely related to sequence-dependent biases in the fragmentation process that can induce preferential cutting sites in multiple copies of the same transcript, thus invalidating the assumptions of unbiased random fragmentation of our theory. While considering many transcripts with different sequences, such biases average out, leading to the previously observed agreement on the average. The second observed effect may be related to RNA degradation from 3'-end in the cell: at the moment of RNA extraction, multiple copies of the same transcript that have been engaged in degradation are likely to have different lengths and thus their 3'-ends at different but neighbouring genomic locations. Such a mechanism will blur the pattern described above and, by shortening RNA molecules, will shift the predictions upstream. Due to those shortcomings, a more detailed model taking fragmentation biases and potential RNA degradation into account will be necessary to provide reliable predictions of the location of the 3'-end of a transcript.  \\

\begin{figure}
\centering
\includegraphics[width=.7\textwidth,trim=2cm 0cm 1cm 0cm, clip=true]{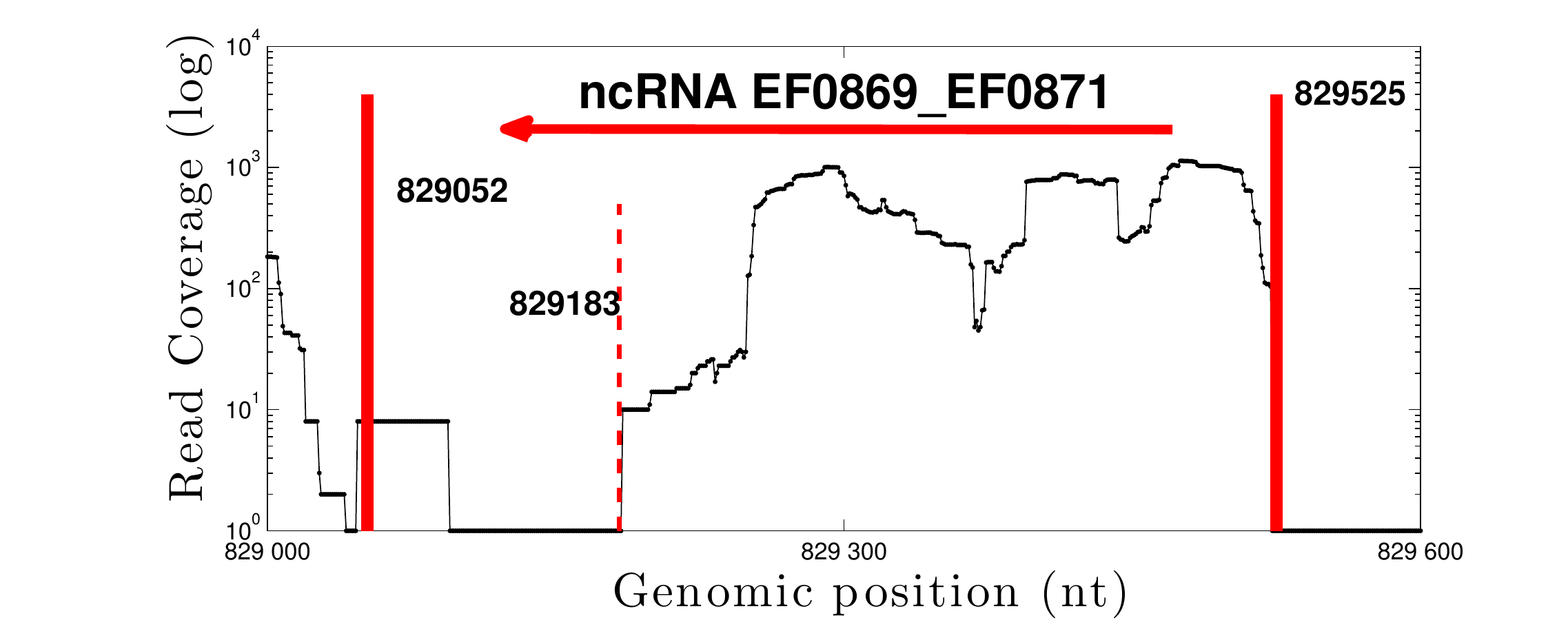}
\caption{Visualisation of the read coverage from the S3sr0A dataset in a region containing a ncRNA detected using microarray \cite{Shioya:2011fk}. The region of the ncRNA is marked with vertical bars. While both methods are in perfect agreement for the 5'-end, a naïve interpretation of the coverage from RNA-seq would place the 3'-end of the transcript at the position indicated by the dashed line, well before the end of the transcript as measured using microarray. \label{fig:samplecomparisonncRNA}}
\end{figure}

A direct consequence of the mechanisms described in this work is that SOLiD single strand RNA-seq cannot be used for accurate determination of 3'-end of transcripts. The potential truncation of transcripts ends can cause problems while comparing results from RNA-seq to other measurement methods such as Northern blot, 3'RACE or microarray (example in \fref{fig:samplecomparisonncRNA}). Furthermore, RNA transcripts longer than the read length (here 50 nt) but significantly shorter than the targeted average length (here 100-120 nt) may simply be totally absent from the final results while one would naïvely expect them to be well visible. 

\section{Conclusions}
\label{sec:conclusion}

We demonstrated that a simple theory based on Kolmogorov's broken stick model explains well the type of signal observed on the average near transcripts 3'-ends in high throughput sequencing experiments.  

The use of this theory for prediction of the 3'-end of individual transcripts has shown only minor improvements compared to a naïve approach. The main reason for this is the presence of clear preferential cleavage sites on transcripts that break the assumption of unbiased fragmentation, essential in our theory. 

Most importantly, we have demonstrated that drawing conclusions about transcript 3'-ends from single stranded RNA-seq experiments on SOLiD must be done with great care. A good understanding of all the mechanisms in the whole RNA-seq pipeline is necessary to give correct interpretation to experimental results.

\ack
Many thanks to Sean P. Kennedy (MetaQuant platform, INRA UMR1319 Micalis) for sequencing the samples of the S55 series as well as his valuable clarifications on the SOLiD protocols. We also thank Aymeric Fouquier d'Hérouël (Institute for Systems Biology) and Francis Repoila (INRA UMR1319 Micalis) for the preparation of RNA samples as well as Ingemar Ernberg (Karolinska Institutet) for his generous hospitality by providing laboratory space and equipment for Aymeric Fouquier d'Hérouël. This work was supported by the Academy of Finland as part of its Finland Distinguished Professor program, project 129024/Aurell, and by the Academy of Finland Center of Excellence COIN. 

\section*{References}
\bibliographystyle{unsrt}
\bibliography{Library}

\end{document}